\documentclass[aps,pre,twocolumn,showpacs,superscriptaddress]{revtex4}
\usepackage{graphicx}

\begin{document}

\title{Emergence of scale-free behavior in networks from limited-horizon linking
and cost trade-offs}

\author{Valmir C. Barbosa}
\affiliation{Programa de Engenharia de Sistemas e Computa\c c\~ao, COPPE,
Universidade Federal do Rio de Janeiro,
Caixa Postal 68511, 21941-972 Rio de Janeiro - RJ, Brazil}

\author{Raul Donangelo}
\affiliation{Instituto de F\'\i sica,
Universidade Federal do Rio de Janeiro,
Caixa Postal 68528, 21941-972 Rio de Janeiro - RJ, Brazil}
\author{Sergio R. Souza}
\affiliation{Instituto de F\'\i sica,
Universidade Federal do Rio de Janeiro,
Caixa Postal 68528, 21941-972 Rio de Janeiro - RJ, Brazil}
\affiliation{Instituto de F\'\i sica,
Universidade Federal do Rio Grande do Sul,
Caixa Postal 15051, 91501-970 Porto Alegre - RS, Brazil}

\begin{abstract}
We study network growth from a fixed set of initially isolated nodes placed at
random on the surface of a sphere. The growth mechanism we use adds edges to the
network depending on strictly local gain and cost criteria. Only nodes that are
not too far apart on the sphere may be considered for being joined by an edge.
Given two such nodes, the joining occurs only if the gain of doing it surpasses
the cost. Our model is based on a multiplicative parameter $\lambda$ that
regulates, in a function of node degrees, the maximum geodesic distance that is
allowed between nodes for them to be considered for joining. For $n$ nodes
distributed uniformly on the sphere, and for $\lambda\sqrt{n}$ within limits
that depend on cost-related parameters, we have found that our growth mechanism
gives rise to power-law distributions of node degree that are invariant for
constant $\lambda\sqrt{n}$. We also study connectivity- and distance-related
properties of the networks.
\end{abstract}


\maketitle

\section{Introduction}

Large-scale networks occurring in a variety of natural, technological, and
social domains have been studied intensely in the last several years. In many
cases, only superficial information is available on the topology of the network
under study, so the most common approach has been to model it as a random graph
\cite{b01} and to describe its properties statistically. Many of these
properties seem to be related to the network's degree distribution, which has
then received considerable attention. In many instances of interest, including
networks related to the Internet or to the WWW, the degree distribution is a
power law. That is, the probability that a randomly chosen node has degree $k$
is proportional to $k^{-\tau}$, in general with $2<\tau<3$. For detailed
information, we refer the reader to the papers collected in \cite{bs03,nbw06}
and to \cite{blmch06}.

An effort closely related to that of characterizing the degree distributions of
existing networks has been to attempt to explain how a power law can emerge from
the underlying mechanisms that govern network evolution. Many of the proposed
explanations have been centered around the so-called Barab\'{a}si-Albert model
\cite{ba99,baj99,dms00,krl00,hk02,br03,cf03,br04}, which is essentially based on
the mechanism, known as preferential attachment, according to which the
appearance of a new edge connecting a new node to a preexisting one is dependent
upon the latter node's current degree in direct proportion. As we argued in an
earlier work \cite{bds06}, preferential attachment often leads to an
unreasonable generative model for networks, since it makes network-growth
decisions depend on global properties and also, for cases like that of computer
networks, implies that node degrees are a more important growth factor than some
cost or efficiency criterion. Other proposals, including our own in
\cite{bds06}, have relied on exclusively local properties
\cite{krrstu00,ccrm02,rs03}.

The mechanism we suggested in \cite{bds06} promotes network growth by the
addition of edges to a fixed set of nodes that, initially, are all part of a
tree. At each time step, two nodes $i$ and $j$ not currently connected by an
edge are randomly selected and an edge is placed between them if a gain function
is found to surpass a cost function for the current network topology. The gain
function seeks to reflect the shortening of distances on the graph that the new
edge may cause between nodes in $i$'s neighborhood and nodes in $j$'s. The cost
function refers to the cost of deploying the connection itself and also to the
cost of possibly having to upgrade $i$ or $j$'s capabilities to accommodate the
new connection. For selected parameter combinations, degree distributions
comprising a two-tier hierarchy of power laws (one for the lower degrees,
another for the higher) are seen to emerge.

One limitation of this mechanism is that any two nodes not currently connected
by an edge may be selected to be the potential end nodes of a new edge. The
trouble with this is not only the implausibility that comes with it in the
context of computer networks, but also the limitation that is indirectly imposed
on the gain functions that can be used. When $i$ and $j$ are selected, the gain
function depends on the current distances between several node pairs, but all we
may assume to be locally available without the need to probe the network beyond
some reasonable depth are upper bounds on these distances. The results we
reported in \cite{bds06} are then based on a gain function that uses such upper
bounds and this is reflected as an oscillatory perturbation in the power laws.

Here we generalize our previous model by dispensing with the need of the initial
spanning tree and also by attaching a geometric reference to each node. We take
each node to be a point on the surface of a sphere and associate with it a
maximum geodesic distance beyond which connections are forbidden. Not only does
this render the model more plausible from the perspective of computer networks,
it also allows our gain function to be expressed in terms of exact distances on
the graph, since these are now obtainable from any node by controlled-depth
incursions into the network.

If the maximum geodesic distance for node interconnection were the same for all
nodes, then the graph obtained by joining all allowed pairs of nodes would be an
instance of what is known as a two-dimensional random geometric graph with
spherical boundary conditions \cite{dc02,p03}. Curiously, for such a graph the
degree distribution is the same Poisson distribution that holds for the
Erd\H{o}s-R\'{e}nyi classic random-graph case \cite{er59}. As we see in the
remainder of the paper, assuming degree-dependent maximum geodesic distances and
deciding whether to interconnect two nodes based on the same trade-offs as in
\cite{bds06} give rise to power-law degree distributions, now without the
oscillations that were caused by using upper bounds on the distances on the
graph. In addition to these distributions, we also study the emergence of a
giant connected component and the relationship between distances on the graph
and geodesic distances.

Our present work is related to the work described in \cite{xs07}, where
constraints similar to ours are imposed on the growing of the networks. However,
the two works remain markedly distinct, since not only the constraints but also
the growth criteria seem to be significantly different in the two cases. In
particular, in \cite{xs07} both minimum and maximum geometric distances are
taken into account (on the plane, which is where the authors assume the network
lies) and, more importantly, a local version of preferential attachment is
used.

\section{The model}

We model network growth by the sequence $G^0,G^1,\ldots$ of undirected graphs,
each on the same set of $n$ nodes, given that in $G^0$ all nodes are isolated
(i.e., $G^0$ has no edges). We assume nodes to lie on the surface of a sphere of
unit radius. For $i$ and $j$ any two nodes, there are two metrics of interest.
One is $\delta_{ij}$, given by the geodesic distance between $i$ and $j$; the
other, for $t=0,1,\ldots$, is $d^t_{ij}$, given by the distance between $i$ and
$j$ in $G^t$ (i.e., the number of edges on the shortest path between $i$ and $j$
in $G^t$).

Let $n^t_i$ be the degree of node $i$ (its number of neighbors) in $G^t$ and
$\rho^t_i$ the geodesic distance beyond which no node may be connected to $i$ in
$G^t$. We use
\begin{equation}
\rho^t_i=\lambda\ln(e+n^t_i)
\label{eq:rho}
\end{equation}
throughout, where $\lambda$ is a parameter, indicating that $\rho^t_i=\lambda$
for isolated nodes (regardless of $t$) and that $\rho^t_i$ increases
logarithmically with $n^t_i$ from then on. In order for $\rho^t_i$ to exclude
no node from the possibility of being connected to $i$ for any $t$, it suffices
to set $\lambda=\pi$.

For $i$ and $j$ any non-neighboring nodes of $G^t$, we define $N^t_i(j)$ to be
the set comprising every neighbor $k$ of $i$ in $G^t$ for which $d^t_{jk}>2$.
Clearly, any node in this set benefits from the addition of an edge between $i$
and $j$, in terms of acquiring a shorter path (of length $2$) to $j$. Not only
this, but we also know that, if $k\in N^t_i(j)$, then
\begin{equation}
\max\{3,d^t_{ij}-1\}\le d^t_{jk}\le d^t_{ij}+1
\label{eq:ineqs1}
\end{equation}
(or else $d^t_{ij}$ could not be the distance between $i$ and $j$ in $G^t$). We
also define $N^t_{ij}$ to be the set of all unordered pairs $(k,l)$ such that
either $k$ or $l$ is a neighbor of $i$, the other node in the pair is a neighbor
of $j$, and moreover $d^t_{kl}>3$. As before, any node pair in this set acquires
a shorter path (of length $3$) between them as a result of adding an edge
between $i$ and $j$. Additionally, for $(k,l)\in N^t_{ij}$,
\begin{equation}
\max\{4,d^t_{ij}-2\}\le d^t_{kl}\le d^t_{ij}+2.
\label{eq:ineqs2}
\end{equation}
[Note that both the inequality pair in (\ref{eq:ineqs1}) and the one in
(\ref{eq:ineqs2}) define nonempty intervals, since $d^t_{ij}>1$.]

For $t\ge 0$, $G^{t+1}$ is obtained from $G^t$ by randomly selecting two nodes,
say $i$ and $j$, such that $d^t_{ij}>1$ and adding an edge between them if the
gain from doing so surpasses the cost, provided
\begin{equation}
\delta_{ij}\le\min\{\rho^t_i,\rho^t_j\}.
\label{eq:maxdist}
\end{equation}
That is, we require not only a positive gain-cost trade-off, but also that the
surface of the spherical cap centered on $i$ include $j$ and the one on $j$
include $i$ (in other words, the least connected node of the pair is the one
that actually determines whether the connection is possible). If either
condition does not hold, then $G^{t+1}=G^t$.

It is important to note that, even though by Eq.~(\ref{eq:rho}) the limitation
on the horizon of admissible connections to a node is expected to grow (albeit
weakly) with the node's degree, no such trend exists regarding the probability
that the node's degree is in fact increased. Instead, it is only the size of the
region inside which a new link to that node may be created that increases. Thus,
clearly, our policy for adding new edges to the graph is not just a more local,
weaker version of preferential attachment, but a wholly different concept.

The gain we use is denoted by $g^t_{ij}$ and gives the total number of edges by
which certain paths become shorter when $i$ and $j$ become neighbors. These
paths are the following: the one between $i$ and $j$, the one between each
$k\in N^t_i(j)$ and $j$, the one between each $k\in N^t_j(i)$ and $i$, and
finally the one between each pair $(k,l)\in N^t_{ij}$. We then have
\begin{eqnarray}
\lefteqn{
g^t_{ij}
=d^t_{ij}-1
+\sum_{k\in N^t_i(j)}(d^t_{jk}-2)}\hspace{0.25in}\nonumber\\
&&+\sum_{k\in N^t_j(i)}(d^t_{ik}-2)
+\sum_{(k,l)\in N^t_{ij}}(d^t_{kl}-3),
\label{eq:gain}
\end{eqnarray}
and it follows from our preceding discussion that
\begin{equation}
g^t_{ij}
\ge 1+\vert N^t_i(j)\vert+\vert N^t_j(i)\vert+\vert N^t_{ij}\vert,
\end{equation}
where $\vert X\vert$ denotes the number of elements of set $X$. Also, it is
important to note that, by (\ref{eq:maxdist}), $g^t_{ij}$ can be obtained by
probing $G^t$ from $i$ nearly exclusively on the surface of $i$'s spherical cap
(some of $j$'s neighbors may be excepted), and conversely from $j$. If nodes are
distributed uniformly on the sphere's surface, then by Eq.~(\ref{eq:rho}) this
amounts to saying that, for fixed $\lambda$, probing $G^t$ from any node is
expected to go no deeper than a distance on the graph that grows only
logarithmically with the node's degree.

The cost component of the trade-off is denoted by $c^t_{ij}$ and given exactly
as in \cite{bds06}, where the reader is referred to for complete details. The
cost $c^t_{ij}$ is given simply by
\begin{equation}
c^t_{ij}=C+D[(n^t_i)^\gamma+(n^t_j)^\gamma],
\label{eq:cost}
\end{equation}
where $C$ is the fixed cost of actually deploying the connection between $i$ and
$j$ and the second term, weighted by the proportionality constant $D$, refers to
amortizing the cost to upgrade the connection capabilities of $i$ and $j$.

\section{Computational results and discussion}

Our results are based on computer simulations for which the parameter values
needed in the cost function of Eq.~(\ref{eq:cost}) are $C=100$, $D=0.1$, and
$\gamma=0.9$. These were the values identified in \cite{bds06} as giving rise
to interesting scale-free behavior and here we use them exclusively. Each
simulation is run through $t=3n^2$ from a randomly chosen $G^0$ instance, which
in turn is obtained by placing the $n$ nodes on the unit-radius sphere uniformly
at random. Most of the data we show are averages over at least $500$ independent
runs (exceptions are the $n=2\,000$ cases with $\lambda\ge 0.4$, for which the
number of runs is between $150$ and $500$).

Whenever necessary for use in the computation of the gain function of
Eq.~(\ref{eq:gain}), distances on the current graph are obtained by searching it
breadth-first from one of the two nodes involved. If $G^t$ is the graph in
question and it has $m^t$ edges, then completing such a search requires
$O(m^t)$ time in the worst case \cite{clrs01}. As we mentioned earlier, however,
in our present case the search need not run over all of $G^t$. In fact, all
necessary distances are expected to be found without the search proceeding any
deeper in $G^t$ than a number of edges that depends only on $\lambda$ and a
slow-growing function of node degrees, since nodes are distributed uniformly on
the sphere.

\begin{figure}
\centering
\includegraphics[scale=0.31]{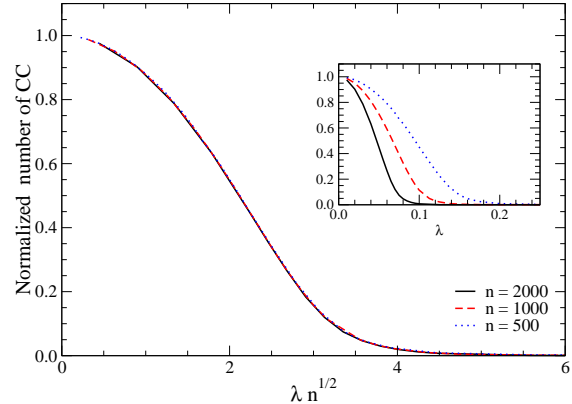}
\caption{(Color online) Normalized number of connected components (CC) as a
function of $\lambda$ (inset) and $\lambda\sqrt{n}$.}
\label{fig:ncc}
\end{figure}

\begin{figure}
\centering
\includegraphics[scale=0.31]{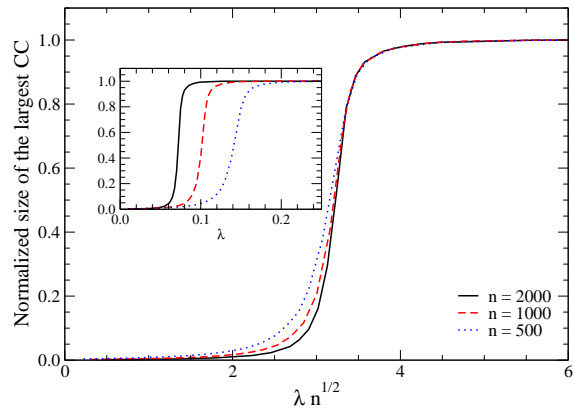}
\caption{(Color online) Normalized size of the largest connected component (CC)
as a function of $\lambda$ (inset) and $\lambda\sqrt{n}$.}
\label{fig:gc}
\end{figure}

\begin{figure}
\centering
\includegraphics[scale=0.31]{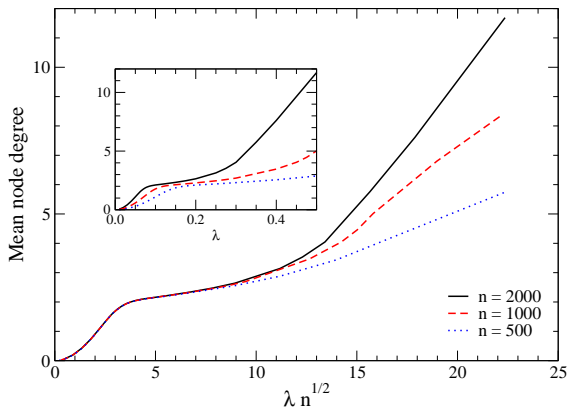}
\caption{(Color online) Mean node degree as a function of $\lambda$ (inset) and
$\lambda\sqrt{n}$.}
\label{fig:nedges}
\end{figure}

Figures~\ref{fig:ncc}--\ref{fig:nedges} show the final connectivity properties
of the graph as a function of $\lambda$ (insets). While for very small $\lambda$
nearly every node is a connected component by itself, increasing the value of
$\lambda$ eventually gives rise to the giant connected component, which
encompasses practically all nodes. Characterizing the sharp transition that
takes place between the two extremes is facilitated by a rescaling of the
$\lambda$ parameter. For relatively small $\lambda$, the surface area of the
spherical cap centered at node $i$ can be approximated by the area of the circle
of radius $\rho^t_i$ centered at $i$. Using this approximation, and by
Eq.~(\ref{eq:rho}), keeping $\lambda\sqrt{n}$ fixed as both $\lambda$ and $n$
vary implies that the expected number of nodes on the spherical cap remains
fixed as well, provided nodes are uniformly distributed on the sphere. Rescaled
plots for the connectivity-related quantities appear in the main plot sets of
Figs.~\ref{fig:ncc}--\ref{fig:nedges}, indicating that the sudden rise of
the giant connected component occurs at $\lambda\sqrt{n}\approx 3$.

The mean node degree increases steadily toward the values above $2$ that
correspond to the presence of the giant connected component \cite{mr95,nsw01},
which already holds for values of $\lambda\sqrt{n}$ around $4$. Above this
value, though, it seems that the existence of this component tends to
discourage the creation of any significant number of additional edges for a
relatively long stretch. As shown in Fig.~\ref{fig:nedges}, only for values
larger than about $11$ are the gain-cost trade-offs once again effective,
allowing for the sustained addition of further edges. It is curious to observe
that the rescaling to $\lambda\sqrt{n}$ breaks down past this value, since the
mean node degree behaves differently for different values of $n$. Part of the
reason why this happens is the approximation of the surface area of the
spherical cap by that of the circle, which becomes inappropriate for large
$\lambda$. We have found, however, that this accounts for only a small fraction
of the observed dependency on $n$. The main reason seems to be related to how
distances on the graph behave as a function of geodesic distances. We return to
this issue shortly.

\begin{figure}
\centering
\includegraphics[scale=0.31]{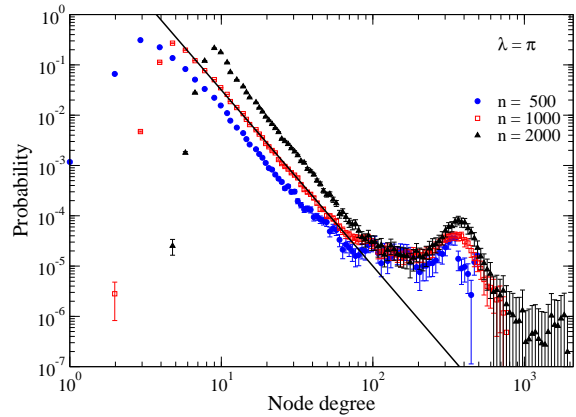}
\caption{(Color online) Node-degree distributions for $\lambda=\pi$. The solid
line gives a power law of exponent $-3.5$. Abscissae are log-binned.}
\label{fig:pkpi}
\end{figure}

\begin{figure}
\centering
\includegraphics[scale=0.31]{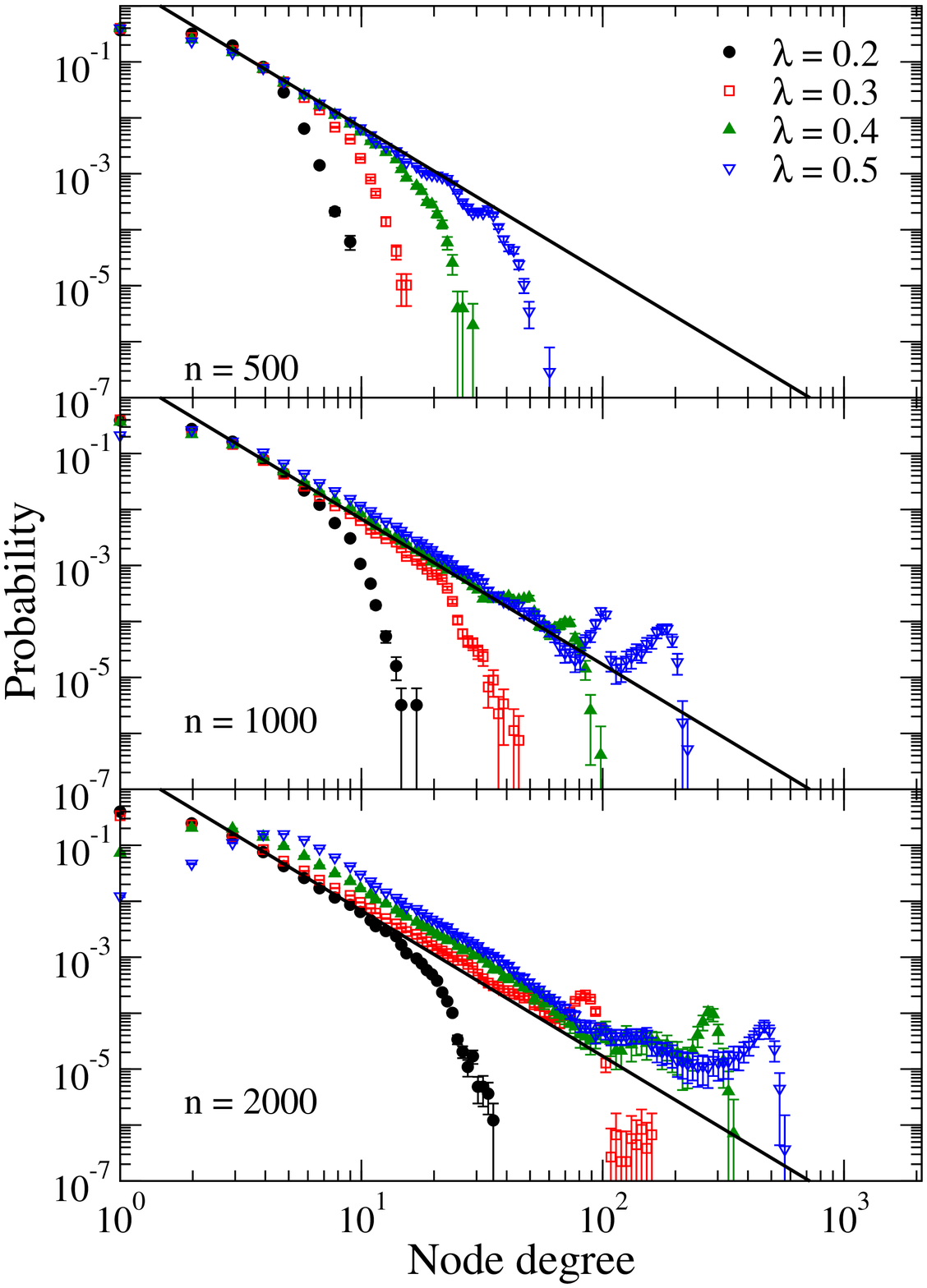}
\caption{(Color online) Node-degree distributions for $\lambda\le 0.5$. The
solid lines give a power law of exponent $-2.6$. Abscissae are log-binned.}
\label{fig:pk}
\end{figure}

\begin{figure}
\centering
\includegraphics[scale=0.31]{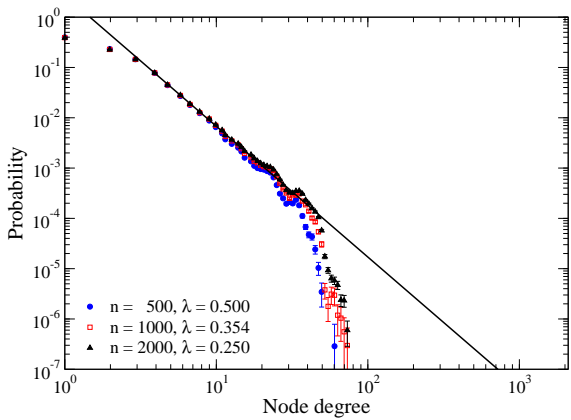}
\caption{(Color online) Node-degree distributions for $\lambda\le 0.5$ such that
$\lambda\sqrt{n}$ is constant. The solid line gives a power law of exponent
$-2.6$. Abscissae are log-binned.}
\label{fig:pkscaling}
\end{figure}

Observed node-degree distributions are given in
Figs.~\ref{fig:pkpi}--\ref{fig:pkscaling}. Figure~\ref{fig:pkpi} is relative
to $\lambda=\pi$ and as such refers back to the unbounded connection
possibilities we explored in \cite{bds06}. For this value of $\lambda$ and the
three values of $n$ used in the figure, the network is well into the regime in
which it almost surely has an all-encompassing connected component
(cf.~Fig.~\ref{fig:gc}), so we expect essentially the same result we obtained
in that earlier occasion. This is in fact what happens, including the power laws
that describe the node-degree distributions up to roughly $100$ and the
transition to a purported, higher-level power law for the higher degrees.

For substantially smaller values of $\lambda$, notice first that, owing to the
uniformity of the node distribution on the sphere, we expect
$(1-\cos\rho^t_i)n/2=n\sin^2(\rho^t_i/2)$ nodes to lie on node $i$'s spherical
cap at time step $t$. This is then an upper bound on the expected value of
$n^t_i$, node $i$'s degree in $G^t$. For fixed $\lambda$, it follows from the
logarithmic dependency in Eq.~(\ref{eq:rho}) that this upper bound grows ever
more slowly as $n^t_i$ increases. The latter, however, is ever less likely,
since by Eq.~(\ref{eq:cost}) the process of increasing $n^t_i$ reflects back on
itself negatively by increasing $c^t_{ij}$ regardless of $j$. It seems, then,
that we are to expect node-degree distributions to undergo a somewhat sharp,
$\lambda$-dependent cutoff at which probability may accumulate.

Node-degree distributions for values of $\lambda$ no greater than $0.5$ are
given in Fig.~\ref{fig:pk}. Notice that a power law does in fact get
established that does not depend on $n$ or $\lambda$, but it is nevertheless
subject to the cutoff mentioned above. The power-law regime lasts for
increasingly larger stretches as $\lambda$ grows but ceases to hold as the
cutoff is approached. For the larger values of $\lambda$ we do observe a certain
accumulation of probability right before this point, but clearly nothing like
the second power law of the $\lambda=\pi$ case seems to be happening now. If, as
before, we now rescale the $\lambda$ parameter so that $\lambda\sqrt{n}$
remains constant as both $\lambda$ an $n$ vary, then by
Fig.~\ref{fig:pkscaling} we see that node degrees are limited by a cutoff
value of about $80$. The plots in this figure are all such that
$\lambda\sqrt{n}\approx 11$, which by Fig.~\ref{fig:nedges} corresponds to the
point beyond which the rescaling is no longer effective. Not only this, but from
Fig.~\ref{fig:nedges} we expect node-degree distributions to change
significantly past this point, eventually approaching the distributions of
Fig.~\ref{fig:pkpi} (the $\lambda=\pi$ case).

\begin{figure}
\centering
\includegraphics[scale=0.31]{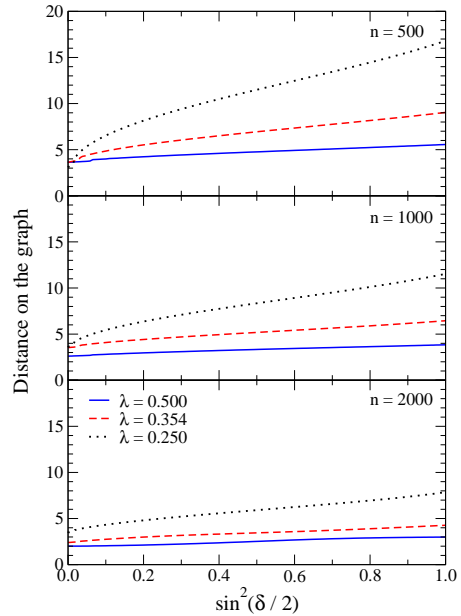}
\caption{(Color online) Average distances on the graph as a function of geodesic
distances for $\lambda\le 0.5$. Abscissae correspond to $100$ fixed-width bins
on the interval $[0,\pi]$.}
\label{fig:sin3}
\end{figure}

One final set of plots is given in Fig.~\ref{fig:sin3} for $\lambda\le 0.5$ to
illustrate the interdependency of distances on the graph and geodesic distances.
Instead of plotting distances on the graph against geodesic distances directly,
we first try to minimize the dependency on $n$ by once again evoking the uniform
distribution of nodes on the sphere. Once we do this, it makes more sense to
rescale each geodesic distance $\delta$ to the surface area of the spherical cap
on which no two nodes are farther apart (geodesic-wise) than $2\delta$. If we
further normalize to the surface area of the entire sphere, then the variable
against which to plot distances on the graph is
$(1-\cos\delta)/2=\sin^2(\delta/2)$. The resulting plots indicate an
approximately linear growth of the distance on the graph, especially for the
larger values of $\lambda$ or $\delta$. The rate of growth is less pronounced
for larger values of $\lambda$ or $n$: in either case, distances on the graph
are expected to be relatively insensitive to the geodesic distances.

In order to understand these trends, consider any two nodes $i$ and $j$ and
recall that $\delta_{ij}$ is the geodesic distance between them. Depending on
$\lambda$, the chance these two nodes constitute a feasible pair in terms of
(\ref{eq:maxdist}) varies considerably. When (\ref{eq:maxdist}) does hold for
some $t$, which by Eq.~(\ref{eq:rho}) tends to happen more often for the larger
values of $\lambda$ or the larger values of $n$, establishing a direct
connection between $i$ and $j$ once the two nodes have been selected depends
exclusively on the gain-cost trade-off, which is a function of the current
graph's topology and independent of $\delta_{ij}$. When (\ref{eq:maxdist}) does
not hold, and this is more often observed for the smaller values of $\lambda$ or
the smaller values of $n$, nodes $i$ and $j$ do not become directly
interconnected just then and the distance between them on the graph tends to be
larger for larger $\delta_{ij}$.

Figure~\ref{fig:sin3} also helps complete our understanding of why the behavior
of the mean node degree depends on $n$ for $\lambda\sqrt{n}>11$ in
Fig.~\ref{fig:nedges}. Recall first that the motivation for the rescaling to
$\lambda\sqrt{n}$ has been to keep the expected number of nodes on a node's
spherical cap fixed as both $\lambda$ and $n$ vary, and therefore induce an
invariant expected behavior as far as the gain-cost trade-offs are concerned.
What Fig.~\ref{fig:sin3} indicates, however, is that distances on the graph
depend on geodesic distances in different ways for fixed $\lambda\sqrt{n}$, and
by Eq.~(\ref{eq:gain}) so do gains, since they depend crucially on distances on
the graph. Our trade-offs can then be expected to operate differently for
different values of $n$ even for fixed $\lambda\sqrt{n}$, indicating that in
general it is not a spherical cap's expected number of nodes that regulates
their operation. But Fig.~\ref{fig:sin3} also suggests that this effect is
relatively negligible when $\sin^2(\delta/2)$ is very small. In fact, for all
values of $\lambda\sqrt{n}$ up to about $11$, this is what happens in
Fig.~\ref{fig:nedges}: in this range, we have mean node degrees of up to about
$3$, and by Eq.~(\ref{eq:rho}) we expect to consider node pairs for joining that
are no farther apart than some geodesic distance $\delta$ such that
$\sin^2(\delta/2)<0.18$. This explains why the rescaling is effective in this
region.

\section{Conclusion}

We have studied networks that grow from a fixed set of isolated nodes initially
placed on the surface of a unit-radius sphere. Growth is promoted by considering
node pairs that are not too far apart on the sphere and weighing, for each pair,
a gain against a cost related to adding an edge between the two nodes. The edge
is added to the network if the gain surpasses the cost. Our study has touched
several relevant issues, such as the connectivity properties of the graph into
which the network settles, the node-degree distribution of this graph, and also
the relationship that exists between distances on the graph and geodesic
distances.

Our mechanism of network growth depends crucially on the parameter, here called
$\lambda$, that regulates, for each node and in a function of its current
degree, the maximum geodesic distance to any node to which it may be connected
by an edge in the graph. Relatively low values of $\lambda$ are attractive
because they lend an additional degree of plausibility to the growth mechanism
by letting nodes depend exclusively on locally available information to decide
whether to get joined by an edge. We have found, by means of computer
simulations, that node degrees at such values of $\lambda$ are distributed
according to a fixed-exponent power law, provided $\lambda$ and $n$, the number
of nodes, are such that $\lambda\sqrt{n}$ remains constant and bounded. For the
cost parameters we used, the required bound on $\lambda\sqrt{n}$ is about $11$,
and the power-law exponent is $-2.6$. This rescaling of $\lambda$ to
$\lambda\sqrt{n}$ is based on the assumption that nodes are distributed
uniformly over the sphere, and we have found it to provide a certain degree of
independence of $n$ in questions related to network connectivity as well.

This paper's study generalizes and extends our own previous work of
\cite{bds06}, respectively by using geometric coordinates to limit the reach of
new edges as they enter the network, and by addressing issues other than that of
node-degree distributions. What the two works have in common is the use of
closely related gain-cost trade-offs as the main drives of network growth. The
fact that, in essence, scale-free properties appear in both models seems to
confirm that trade-offs such as the ones we have used have an important role to
play in the evolution of real-world computer networks.

\begin{acknowledgments}
We acknowledge partial support from CNPq, CAPES, FAPERJ BBP grants, and the
joint PRONEX initiative of CNPq and FAPERJ under contracts 26.171.176.2003 and
26.171.528.2006.
\end{acknowledgments}

\bibliography{gnet}
\bibliographystyle{apsrev}

\end{document}